\documentclass{jpsj2}
\usepackage{bm,graphicx}
\title{Statistical Mechanics of On-line Ensemble Teacher Learning through a Novel Perceptron Learning Rule}
\author{\textsc{Kazuyuki Hara}\thanks{E-mail address: hara.kazuyuki@nihon-u.ac.jp} and \textsc{Seiji Miyoshi}$^{1}$
}
\inst{College of Industrial Technology, Nihon University, \\
1-2-1 Izumi-cho, Narashino, Chiba 275-8575, Japan \\
$^{1}$Faculty of Engineering Science, Kansai University, \\ 
3-3-35 Yamate-cho,  Suita, Osaka 564-8680, Japan}

\abst{
In ensemble teacher learning, ensemble teachers have only uncertain information about the true teacher, and  this information is given by an ensemble consisting of an infinite number of ensemble teachers whose variety is sufficiently rich.  In this learning, a student learns from an ensemble teacher that is iteratively selected randomly from a pool of many ensemble teachers. 
An interesting point of ensemble teacher learning is the asymptotic behavior of the student to approach the true teacher by learning from ensemble teachers.   The student performance is improved by using the Hebbian learning rule in the learning. However, the perceptron learning rule cannot improve the student performance. On the other hand, we proposed a perceptron learning rule with a margin.  This learning rule is identical to the perceptron learning rule when the margin is zero and identical to the Hebbian learning rule when the margin is infinity. Thus, this rule connects the perceptron learning rule and the Hebbian learning rule continuously through the size of the margin. Using this rule, we study changes in the learning behavior from the perceptron learning rule to the Hebbian learning rule by considering several margin sizes. From the results, we show that by setting a margin of $\kappa > 0$, the effect of an ensemble appears and becomes significant when a larger margin $\kappa$ is used. }

\kword{ensemble teacher learning, perceptron learning rule, margin, on-line learning, generalization error, statistical mechanics}

\begin{document}
\maketitle
\section{Introduction}
Ensemble learning using an ensemble of many weak learners (referred to as students) improves the performance of a learning machine. The ensemble is obtained by calculating the average student output. 
Bagging\cite{Breiman1996} and boosting\cite{Freund1997} are each a type of ensemble learning. 
Ensemble learning is classified with respect to how a new student is added and how students are combined. In particular, when the students are nonlinear perceptrons, the space spanned by combining students differs from the original space; thus, ensemble learning improves the learning machine performance. \cite{Murata2004}. 

Miyoshi and Okada proposed ensemble teacher learning as an alternative rule to ensemble learning\cite{Miyoshi2006}. This rule employs a true teacher, ensemble teachers, and a student. They assume that the ensemble teachers have uncertain information about the true teacher and that this information is given by an ensemble consisting of an infinite number of ensemble teachers, where the variety of ensemble teachers is sufficiently rich.
An interesting point of this rule is the asymptotic behavior of the student to approach the true teacher by learning from ensemble teachers. 
The key point of this rule is that even if the student learns from an ensemble teacher selected randomly from a pool of many ensemble teachers, the student learns from the ensemble of teachers as a result\cite{Okada}. This is similar to the bagging\cite{Breiman1996} or the boosting\cite{Freund1997} using the ensemble of many students. 
Utsumi et al.\cite{Utsumi2007} showed that the student performs better than the ensemble teachers after  learning when the Hebbian learning rule\cite{Hebb} is used. However, the perceptron learning rule\cite{Minsky1969} cannot improve the student performance. 
They did not give a  reason for this failure, which is an open problem.   

On the other hand, we proposed a perceptron learning rule with a margin.\cite{Hara2004}.
This rule is identical to the perceptron learning rule when the margin is zero and  identical to the Hebbian learning rule when the margin is infinity. Otherwise, it lies somewhere between these two rules. Therefore, we considered that by using the perceptron learning rule with a margin in ensemble teacher learning, we can investigate the change in the learning behavior from the perceptron learning rule to the Hebbian learning rule by changing the size of the margin to solve the above open problem\cite{Hara2011} .

In this paper, we first show that in ensemble teacher learning, the student does not learn from some ensemble teachers in the case of the perceptron learning rule, and this is the cause of the failure of ensemble teacher learning.  Then we show how the margin controls the learnable regions of the student and that a small margin allows the student to learn from more ensemble teachers than when using a zero margin. 
Next, we provide a theoretical analysis of the proposed rule through the derivation of coupled differential equations that  depict the learning behavior using statistical mechanics methods\cite{nishimori2001,Saadbook,Hara2009}. After that, we solve the order parameter equations numerically and show the behavior of the generalization error. Finally, we show the validity of the proposed rule. 

\section{Model}

\begin{figure}[t]
\begin{center}
\includegraphics[width=5cm]{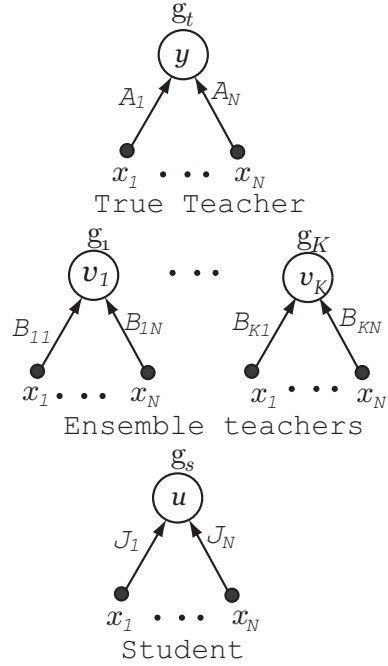}
\end{center}
\caption{\label{arch}Architecture of true teacher, $K$ ensemble teachers, and a student.}
\end{figure}

In this paper, we consider a true teacher, $K$ ensemble teachers, and a student. These are all nonlinear simple perceptrons as shown in Fig. \ref{arch}.

We first formulate the architecture of the perceptrons.  As shown in the figure, all perceptrons have $N$ input and one output, and the true teacher, ensemble teachers, and the student are with connection weights $\bm{A}=(A_1, \ldots , A_N)$, $\bm{B}_k=(B_{k1},\ldots ,B_{kN})$, and $\bm{J}(m)=(J_1(m),\ldots ,J_N(m))$, respectively. Here, $k=1, \ldots ,K$ and $m$ denots learning iterations. In the figure, learning iteration $m$ is ignored. For simplicity, the connection weights of the true teacher, ensemble teachers, 
and student are simply called the true teacher, ensemble teachers, and student, respectively.
We assume that the true teacher, ensemble teachers, and student receive an $N$-dimensional input 
$\bm{x}(m) = (x_1(m) , \ldots , x_N(m))$ at the $m$th learning iteration, and they output $\mbox{g}_t(m)=\mbox{sgn}(y(m))$, $\mbox{g}_k(m)=\mbox{sgn}(v_k(m)), k=1, \ldots , K$, and $\mbox{g}_s(m)=\mbox{sgn}(u(m))$, respectively.  Here, $\mbox{sgn}(z)=1$ when $z \ge 0$, and $-1$ otherwise. 
The inner potentials of the true teacher $y(m)$, ensemble teachers $v_k(m)$, and student $u(m)$ are
 
\begin{align}
y(m)=\sum_{i=1}^N A_i x_i(m) = \bm{A}\cdot \bm{x}(m),\\
v_k(m)=\sum_{i=1}^N B_{ki} x_i(m)=\bm{B}\cdot \bm{x}(m),\\
u(m)=\sum_{i=1}^N J_i(m) x_i(m)=\bm{J}(m)\cdot \bm{x}(m).
\end{align}

Next, we formulate the setting of the perceptrons for theoretical analysis.
We assume that the elements $x_i(m)$ of the independently drawn input $\bm{x}(m)$ are uncorrelated Gaussian random variables with zero mean and a variance of $1/N$; that is, the $i$th element of the input is drawn from a probability distribution $P(x_i)$. The statistics of the input $\bm{x}(m)$ at the limit of $N \rightarrow \infty$ are 

\begin{equation}
\left<x_i(m)\right>=0, \left<(x_i(m))^2\right>=\frac{1}{N}
\end{equation}

\noindent
where $\left< \cdot \right>$ denotes a mean. 
We assume that each element of the true teacher $A_i$ and those of the initial student $J_i(0)$ are drawn from a Gaussian distribution of zero mean and unit variance. Here, $i=1, \ldots, N$. 
Some elements $B_{ki}$ are equal to $A_i$ multiplied by $-1$, and the others are equal to $A_i$.
The elements of $B_{ki}$ that are equal to $-A_i$ are independent of the value of $A_i$.
Hence, $B_{ki}$ also obeys a Gaussian distribution of zero mean and unit variance.
The statistics of the true teacher $\bm{A}$, ensemble teachers $\bm{B}_k$, and  initial student $\bm{J}(0)$ at the limit are 

\begin{align}
\left<A_i\right>=0, \left<(A_i)^2\right>=1, \\
\left<B_{ki}\right>=0, \left<(B_{ki})^2\right>=1, \\
\left<J_i(0)\right>=0, \left<(J_i(0))^2\right>=1.
\end{align}

\noindent
The direction cosine between $\bm{J}$ and $\bm{A}$ is $R_J$, that between $\bm{A}$ and $\bm{B}_k$ is $R_{Bk}$, that between $\bm{J}$ and $\bm{B}_k$ is $R_{BkJ}$, and that between $\bm{B}_k$ and $\bm{B}_{k^{\prime}}$ is $q_{kk^{\prime}}$. These are the order parameters of the learning system and are defined as

\begin{align}
R_{J}=\frac{\bm{A}\cdot \bm{J}}{\|\bm{A}\| \| \bm{J}\|}, \\
R_{Bk}=\frac{\bm{A}\cdot \bm{B}_k}{\|\bm{A}\| \|\bm{B}_k\|}, \\
R_{BkJ}=\frac{\bm{B}_k \cdot \bm{J}}{\|\bm{B}_k\| \|\bm{J}\|}, \\
q_{kk'}=\frac{\bm{B}_k \cdot \bm{B}_{k'}}{\|\bm{B}_k\| \| \bm{B}_{k'}\|}.
\end{align}

\begin{figure}
\includegraphics[width=8cm]{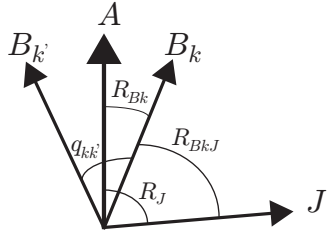}
\caption{\label{direction_cosine}True teacher $\bm{A}$, ensemble teacher $\bm{B}_k$, and student $\bm{J}$. $q_{kk'}$, $R_{J}$, $R_{Bk}$, and $R_{BkJ}$ are direction cosines.}
\end{figure}

\noindent
Figure \ref{direction_cosine} depicts the relationship between the true teacher $\bm{A}$, an ensemble teacher $\bm{B}_k$, the student $\bm{J}$, and the direction cosines $R_{J}$, $R_{Bk}$, $R_{BkJ}$, and $q_{kk'}$.

We assume the thermodynamic limit of $N \rightarrow \infty$. Therefore, 

\begin{equation}
\| \bm{x} \|=1
\end{equation}

\noindent
and
\begin{align}
\|\bm{A}\|&=\sqrt{N}, \|\bm{B}_k\|=\sqrt{N}, \|\bm{J}(0)\|=\sqrt{N}.
\end{align}

\noindent
Generally, the norm of the student $\|\bm{J}(m)\|$ changes as the time step proceeds. Therefore, the ratio $l$ of the norm to $\sqrt{N}$ is considered and is called the length of the student  $\bm{J}$. We define the length of the student as 

\begin{equation}
l(m) = \frac{1}{\sqrt{N}} \| \bm{J}(m)\|
\end{equation}

\noindent
At the thermodynamic limit of $N \rightarrow \infty$, the distribution of the input potential of the true teacher $P(y)$, those of the ensemble teachers $P(v_k)$, and that of the student $P(u)$ follow a Gaussian distribution with zero mean and  covariance matrix $\Sigma_k$: 

\begin{align}
P(y,v_k,u)&=\frac{1}{\left(2\pi\right)^{\frac{3}{2}} | \Sigma_k |^{\frac{1}{2}}}\exp\left(-\frac{(y,v_k,u) \Sigma_k^{-1} (y,v_k,u)^T}{2}\right)\\
\Sigma_k&=\left( 
  \begin{array}{ccc}
     1 & R_{Bk} & R_{J} \\
     R_{Bk} & 1 & R_{BkJ} \\
     R_{J} & R_{BkJ} & l^2\\
   \end{array} \right) .
\end{align}

Next, we define the generalization error $\epsilon_g$. It is given by the student error $\epsilon(m)=\Theta(-y(m)\cdot u(m))$ averaged over all possible inputs. Here the learning iteration $m$ is omitted for simplicity.  Here, $\Theta(z)=1$ when $z \ge 0$, and $0$ otherwise. 

\begin{equation}
\epsilon_g=\int d\bm{x}P(\bm{x})\epsilon=\int dyduP(y,u)\Theta(-y\cdot u)=\frac{1}{\pi}\arctan\frac{\sqrt{1-R_J^2}}{R_J} \label{generalization_error}
\end{equation}

\noindent
$P(y,u)$ is the joint distribution of $y$ and $u$.

\subsection{Ensemble teacher learning through the perceptron learning rule}

We next introduce ensemble teacher learning\cite{Miyoshi2006}. 
This learning uses a true teacher, ensemble teachers, and a student. 
The student learns from an ensemble teacher that is randomly selected from a pool of $K$ ensemble teachers. 
The ensemble teachers have only rough information about the true teacher, and this information is given by an ensemble  consisting of  an infinite number of ensemble teachers whose variety is sufficiently rich. 
We assume that time $t$ is defined as $t=m/N$ and $N \rightarrow \infty$; thus, there are $N\Delta t$ iterations in a macroscopic interval of $t \rightarrow t+\Delta t$. In this limit, the difference between the update in each learning (microscopic interval) becomes sufficiently small  in comparison with the size of $\|\bm{J}\|$ to be replaced by the average update over $N$ iterations. Therefore, the student learns an amount equal to the average amount learned from all the ensemble teachers within $\Delta t$\cite{Hara2009}.    This is the mechanism of ensemble teacher learning.

We consider the case when the perceptron learning rule\cite{Minsky1969} is used as a learning rule for ensemble teacher learning. The learning equation is

\begin{equation}
\bm{J}(m+1)=\bm{J}(m)+\eta \Theta \left(-\mbox{sgn}(u(m)) \ \mbox{sgn}(v_{k'(m)})\right) \mbox{sgn}(v_{k'(m)}) \bm{x}(m). \label{ensemble_learning_equation}
\end{equation}

\noindent
Here, subscript $k'(m)$ denotes the ensemble teacher selected at the $m$th iteration, and $\Theta(z)$ is a step function; $\Theta(z)=+1$ when $z > 0$, and $0$ otherwise. 
From eq. (\ref{ensemble_learning_equation}), it can be shown that the student learns from a selected ensemble teacher if the sign of the student output differs from that of the ensemble teacher output. In other words, the student does not learn from some ensemble teachers and ensemble learning does not occur. 

\begin{figure}[hbt]
\begin{minipage}{7.5cm}
\begin{center}
\includegraphics[width=7.5cm]{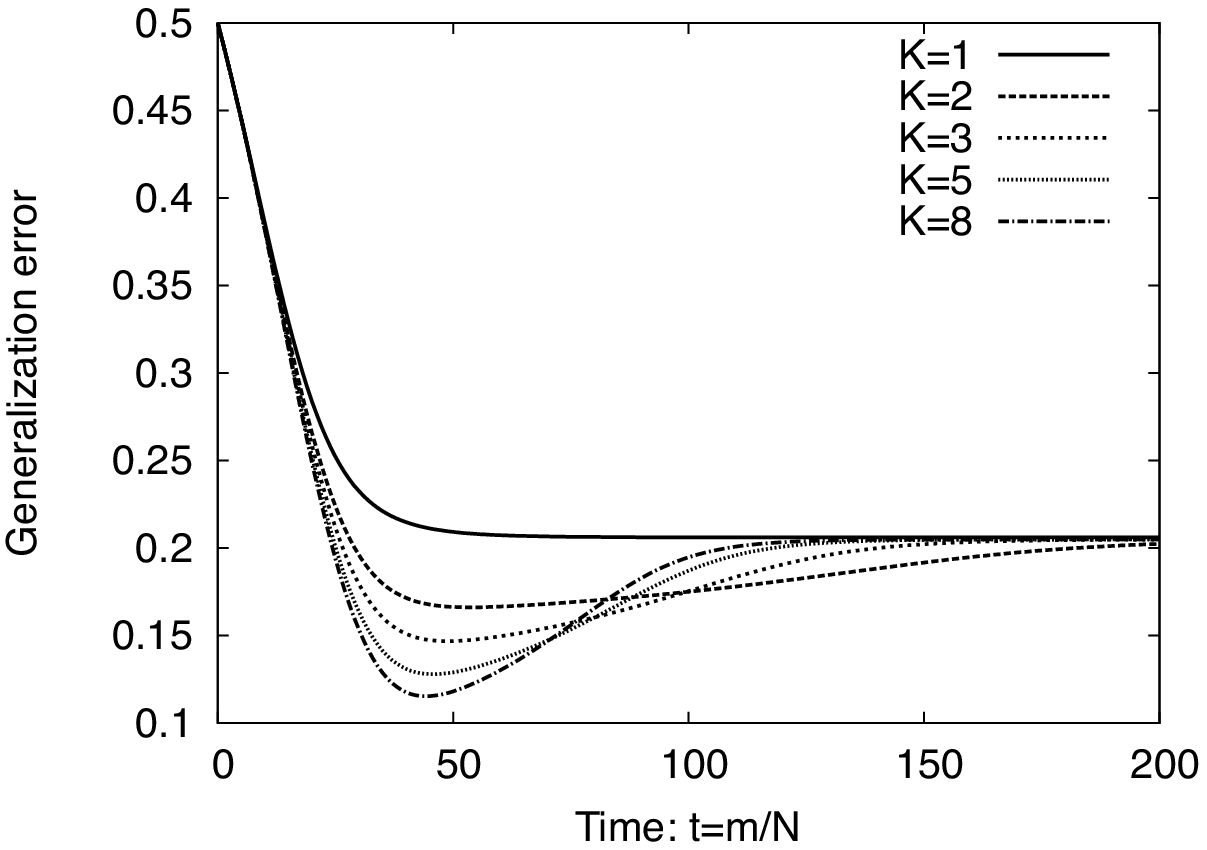}
(a) Generalization error.
\end{center}
\end{minipage}
\begin{minipage}{7.5cm}
\begin{center}
\vspace{-0.0cm}\includegraphics[width=7.2cm]{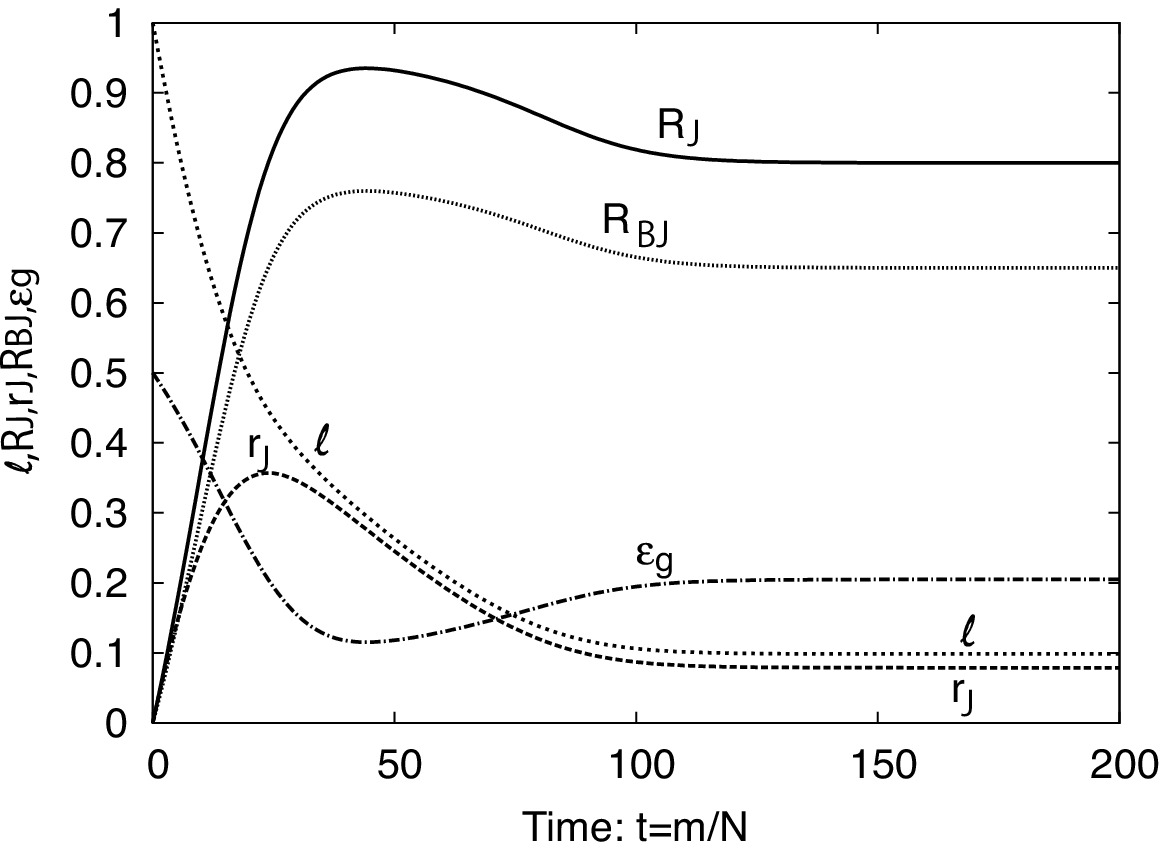}
(b) Order parameters ($K=8$).
\end{center}
\end{minipage}
\caption{\label{pt}Time evolution of (a) generalization error and (b) order parameters of ensemble teacher learning through the perceptron learning rule. The learning step size $\eta$ is set to 0.1. The initial conditions are $R_{B}=0.8$, $R_J(0)=R_{BJ}(0)=0$, $q=0.6$, and $l(0)=1$.}
\end{figure}

Figure \ref{pt}(a) shows the time evolution of the generalization error for the cases of $K=1, 2, 3, 5$, and $8$. 
Figure \ref{pt}(b) shows the time evolution of the order parameters $R_J(t)$, $r_J(t)=R_J l$, $R_{BkJ}(t)$, $l(t)$, and $\epsilon_g(t)$ for the case of $K=8$.  Here we assume that all $R_{BkJ}$ and $R_{Bk}$ are independent of the subscript $k$; thus, we write $R_{BkJ}(t)=R_{BJ}(t)$ and $R_{Bk}(t)=R_B(t)$.
Analytical solutions are used (for the analytical solutions, see the Appendix). The horizontal axis in Figs. \ref{pt}(a) and \ref{pt}(b) is the normalized time $t=m/N$.  The vertical axis in Fig. \ref{pt}(a) is the generalization error and that in Fig. \ref{pt}(b) is the order parameter. In Fig. \ref{pt}(b), the argument $t$ of the order parameters is omitted for simplicity.
In Fig. \ref{pt}, the learning step size is $\eta=0.1$. The initial conditions are $R_{B}=0.8$, $R_{BJ}(0)=R_J(0)=0$, $q=0.6$, and $l(0)=1$.

\noindent
As shown in Fig. \ref{pt}(a), the generalization error decreases with larger $K$ and overshoots in the early stage of learning, but the errors eventually become the same regardless of $K$.    
Therefore, the effect of using many ensemble teachers asymptotically disappears.
This phenomenon can be  explained by considering Fig. \ref{pt}(b) and eq. (\ref{drj}) as follows.
From eq. (\ref{drj}), $r_j(t)$ appears to remain constant when $R_J(t)=R_B$ ($R_J(t)$ is the overlap between the true teacher and student, and $R_B(=R_{Bk}$) is the overlap between the true teacher and ensemble teacher. $R_B$ is a constant value).  However, from Fig. \ref{pt}(b), at the early stage of learning, the student length $l(t)$ changes with time; thus, $R_J(t)$ and $r_j(t)$ change until $l(t)$ remains constant. When $l(t)$ approaches an asymptotic value, $r_j(t)$ also remains constant and $R_J(t)=R_B$ is satisfied. 
This means that $R_J(t)$ cannot exceed $R_B$, which is why  the generalization errors eventually became the same regardless of $K$. 

\section{Theory of  Proposed Ensemble Teacher Learning}

In this section, we propose ensemble teacher learning through the perceptron learning rule with a margin, and we construct a theory supporting this rule.

The cause of the diminishing effect of using many ensemble teachers is that the perceptron learning rule does not allow learning from ensemble teachers who have the same sign of outputs as those of the student. To avoid this problem while using the perceptron learning rule, we introduce the perceptron learning rule with a margin\cite{Hara2004}. The learning equation is

\begin{equation}
\bm{J}(m+1)=\bm{J}(m)+\eta \Theta \left(\kappa -u(m) \ \mbox{sgn}(v_{k'(m)})\right) \mbox{sgn}(v_{k'(m)}) \bm{x}(m).\label{kappa}
\end{equation}

\noindent
Here, $\kappa$ is a positive constant and subscript $k'(m)$ denotes the ensemble teacher selected at the $m$th iteration. As shown in eq. (\ref{kappa}), the student learns from this ensemble teacher if the student output satisfies
 $u(m) < \kappa$, and the learnable region in the input space expands as $\kappa$ increases.
Therefore, in the proposed rule, the student achieves a better performance by learning from more ensemble teachers in a macroscopic interval. 
Note that when $\kappa \rightarrow \infty$, this learning rule is identical to the Hebbian learning rule, and when $\kappa \rightarrow 0$, it is identical to the perceptron learning rule. In other words, the perceptron learning rule with a finite nonzero margin is intermediate between the Hebbian learning rule and the perceptron learning rule. 

Next, we construct the theory of the proposed rule.
We derive differential equations for the order parameters in the proposed rule. 
We introduce closed differential equations for the order parameters that depict the dynamics of the learning system\cite{Utsumi2007}. $r_{BkJ}=R_{BkJ} l$ and $r_{J}=R_{J} l$ are assumed in order to simplify the analysis. 
We omit the learning iteration $m$ here.

\begin{align}
\frac{dr_{BkJ}}{dt}&=\frac{\eta}{K}\sum_{k'=1}^{K}\left<f_{k'}v_k\right> , \label{drBkjdt}\\
\frac{dr_J}{dt}&=\frac{\eta}{K}\sum_{k=1}^K\left<f_k y \right> , \label{drJdt}\\
\frac{dl}{dt}&=\frac{1}{K}\sum_{k=1}^K\left\{\eta \left<f_k\right>+\frac{\eta^2}{2l}\left<f_k^2\right>\right\} . \label{dldt}
\end{align}

\noindent
Here, $f_k=\Theta\left(\kappa-u(m) \ \mbox{sgn}(v_k)\right) \cdot \mbox{sgn}(v_k)$. Then, the  following equations are obtained:

\begin{align}
\left<f_{k'}v_k\right>&=\sqrt{\frac{2}{\pi}}\left [q_{kk'} \mbox{H}\left(\frac{-\frac{\kappa}{l}}{\sqrt{1-R_{Bk'J}^2}}\right) 
-  R_{BkJ} \exp\left(-\frac{\kappa^2}{2l^2}\right)\mbox{H}\left(\frac{-\frac{\kappa}{l}R_{Bk'J}}{\sqrt{1-R_{Bk'J}^2}}\right)\right ], \nonumber \\
\left<f_{k} y\right>&=\sqrt{\frac{2}{\pi}}\left [R_{Bk} \mbox{H}\left(\frac{-\frac{\kappa}{l}}{\sqrt{1-R_{BkJ}^2}}\right) - R_{J} \exp\left(-\frac{\kappa^2}{2l^2}\right)\mbox{H}\left(\frac{-\frac{\kappa}{l}R_{BkJ}}{\sqrt{1-R_{BkJ}^2}}\right)\right ], \nonumber \\
\left<f_{k}u\right>&=\sqrt{\frac{2}{\pi}}\left [R_{BkJ} \mbox{H}\left(\frac{-\frac{\kappa}{l}}{\sqrt{1-R_{BkJ}^2}}\right) - \exp\left(-\frac{\kappa^2}{2l^2}\right)\mbox{H}\left(\frac{-\frac{\kappa}{l}R_{BkJ}}{\sqrt{1-R_{BkJ}^2}}\right)\right ],  \nonumber \\
\left<f_k^2\right>&=2\int_0^{\infty}Du \mbox{H}\left(\frac{R_{BkJ}v_k-\frac{\kappa}{l}}{\sqrt{1-R_{BkJ}^2}}\right). \nonumber
\end{align}

\noindent
Here, 

\begin{equation}
\mbox{H}(x)=\int_x^{\infty}D x = \int_x^{\infty}\frac{dx}{\sqrt{2\pi}}\exp\left(-\frac{x^2}{2}\right).\nonumber
\end{equation}

\noindent
We assume that $R_{BkJ}=R_{BJ}$, $R_{Bk}=R_B$, and $q_{kk'}=q$ when $k\neq k^{\prime}$ and that $q_{kk'}=1$ when $k=k^{\prime}$ in order to simplify the analysis. $K$ ensemble teachers are used. Then we obtain 

\begin{align}
\frac{dr_{BJ}}{dt}=&\eta \sqrt{\frac{2}{\pi}}\left[ \frac{1+(K-1)q}{K} \mbox{H}\left(\frac{-\frac{\kappa}{l}}{\sqrt{1-R_{BJ}^2}}\right)- R_{BJ} \exp\left(-\frac{\kappa^2}{2l^2}\right)\mbox{H}\left(\frac{-\frac{\kappa}{l}R_{BJ}}{\sqrt{1-R_{BJ}^2}}\right)\right], \label{drBJ-pro}\\
\frac{dr_J}{dt}=&\eta \sqrt{\frac{2}{\pi}} \left[R_B \mbox{H}\left(\frac{-\frac{\kappa}{l}}{\sqrt{1-R_{BJ}^2}}\right)-R_{J} \exp\left(-\frac{\kappa^2}{2l^2}\right)\mbox{H}\left(\frac{-\frac{\kappa}{l}R_{BJ}}{\sqrt{1-R_{BJ}^2}}\right)\right], \label{drJ-pro}\\
\frac{dl}{dt}=&\eta \sqrt{\frac{2}{\pi}} \left[R_{BJ} \mbox{H}\left(\frac{-\frac{\kappa}{l}}{\sqrt{1-R_{BJ}^2}}\right)-\exp\left(-\frac{\kappa^2}{2l^2}\right)\mbox{H}\left(\frac{-\frac{\kappa}{l}R_{BJ}}{\sqrt{1-R_{BJ}^2}}\right)\right] \nonumber \\
+& \frac{\eta^2}{l}\int_{0}^{\infty} D v_k \mbox{H}\left(\frac{R_{BJ}-\frac{\kappa}{l}}{\sqrt{1-R_{BJ}^2}}\right). \label{dl-pro}
\end{align}

\section{Results}
\begin{figure}[htb]
\begin{center}
\includegraphics[width=7.5cm]{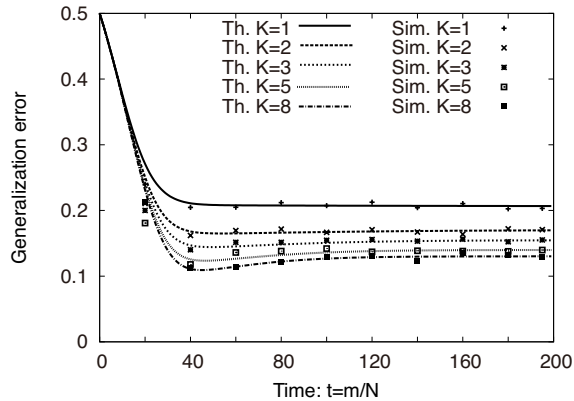}
\end{center}
\caption{\label{proposed}Time evolution of generalization error for proposed rule. The margin $\kappa$ is $0.1$ and the learning rate is $\eta=0.1$.}
\end{figure}

We solved the closed-order parameter equations (eqs. (\ref{drBJ-pro}) to (\ref{dl-pro})) of the proposed rule numerically and then substituted the solutions into eq. (\ref{generalization_error}) to obtain the generalization error. We compared the errors with those of computer simulations. Figure \ref{proposed} shows the time evolution of the generalization error where the number of ensemble teachers was $K=1, 2,3,5,$ or $8$. The horizontal axis is normalized time $t=m/N$, where $m$ is the learning iteration. The vertical axis is the generalization error. The initial conditions are $R_{BJ}(0)=R_J(0)=e^{-10}$ and $l(0)=1$. We set $R_B=0.8$ and $q=0.6$.  The margin $\kappa$ was 0.1, and the learning step size was $\eta=0.1$. Figure \ref{proposed} shows analytical solutions and computer simulation results. The analytical solutions are shown by lines labeled  "Th." followed by the number of outputs $K$. For the computer simulations, $N=1000$ and the results are averaged over 10 trials. These results are shown by marks labeled  "Sim."  followed by the number of outputs $K$. One thousand samples were used to calculate the mean error at each learning iteration. The figure shows that the analytical solutions agreed with those of the computer simulations, confirming the validity of the analytical solutions. 
In Fig. \ref{pt}(a), the effect of the ensemble obtained from many ensemble teachers asymptotically disappeared through the perceptron learning rule. However,  in Fig. \ref{proposed}, the generalization error decreased with increasing $K$ through the perceptron learning rule with a margin; thus, our objective is achieved. 

\begin{figure}[htb]
\begin{minipage}{7.5cm}
\begin{center}
\includegraphics[width=7.5cm]{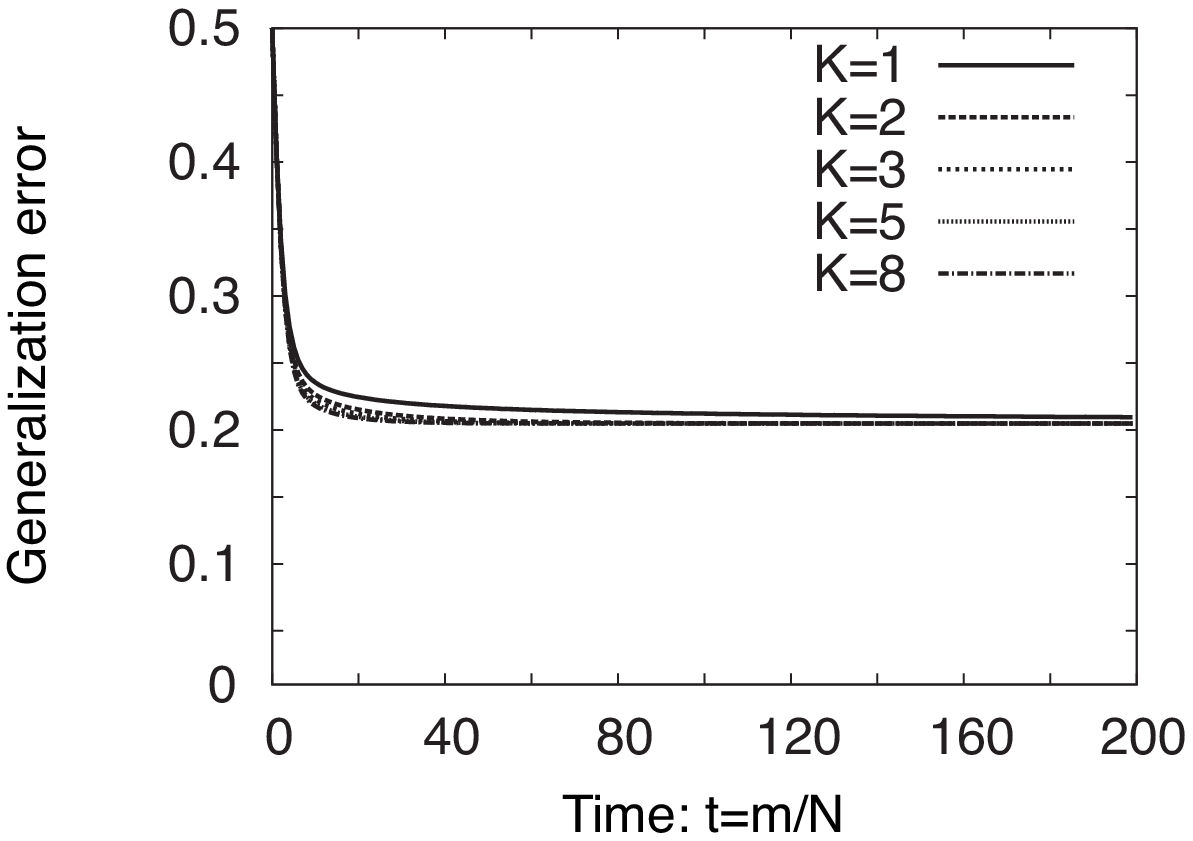}
(a) $\kappa=0$
\end{center}

\end{minipage}
\begin{minipage}{7.5cm}
\begin{center}
\includegraphics[width=7.5cm]{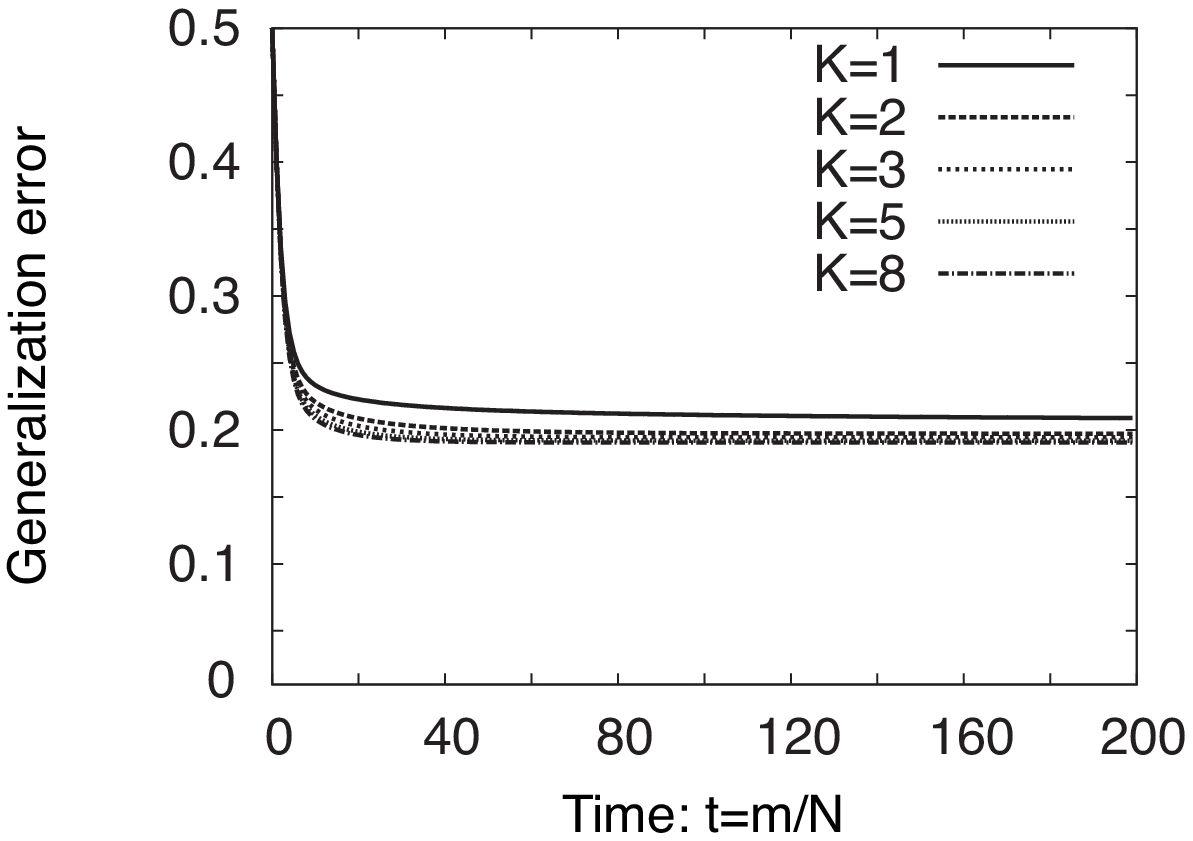}
(b) $\kappa=0.1$
\end{center}

\end{minipage}
\begin{minipage}{7.5cm}
\begin{center}
\includegraphics[width=7.5cm]{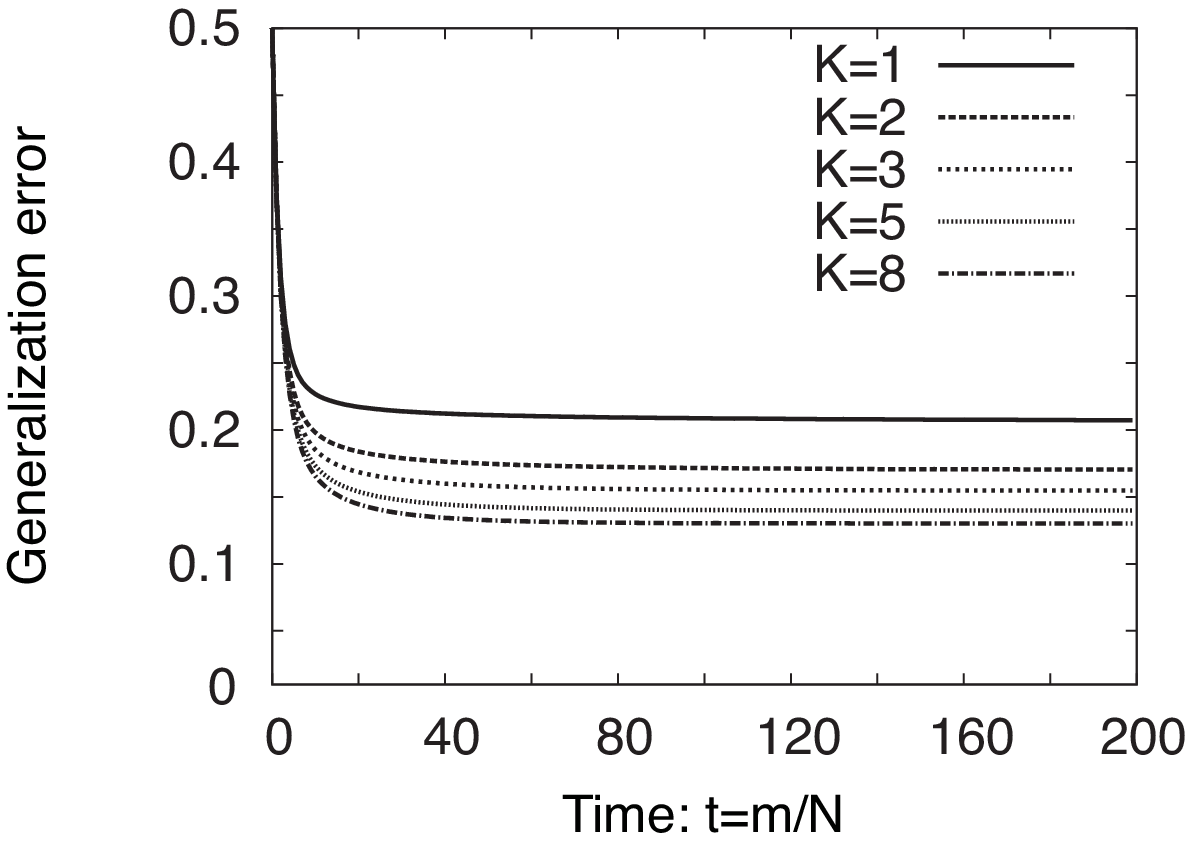}
(c) $\kappa=1$
\end{center}
\end{minipage}
\begin{minipage}{7.5cm}
\begin{center}
\includegraphics[width=7.5cm]{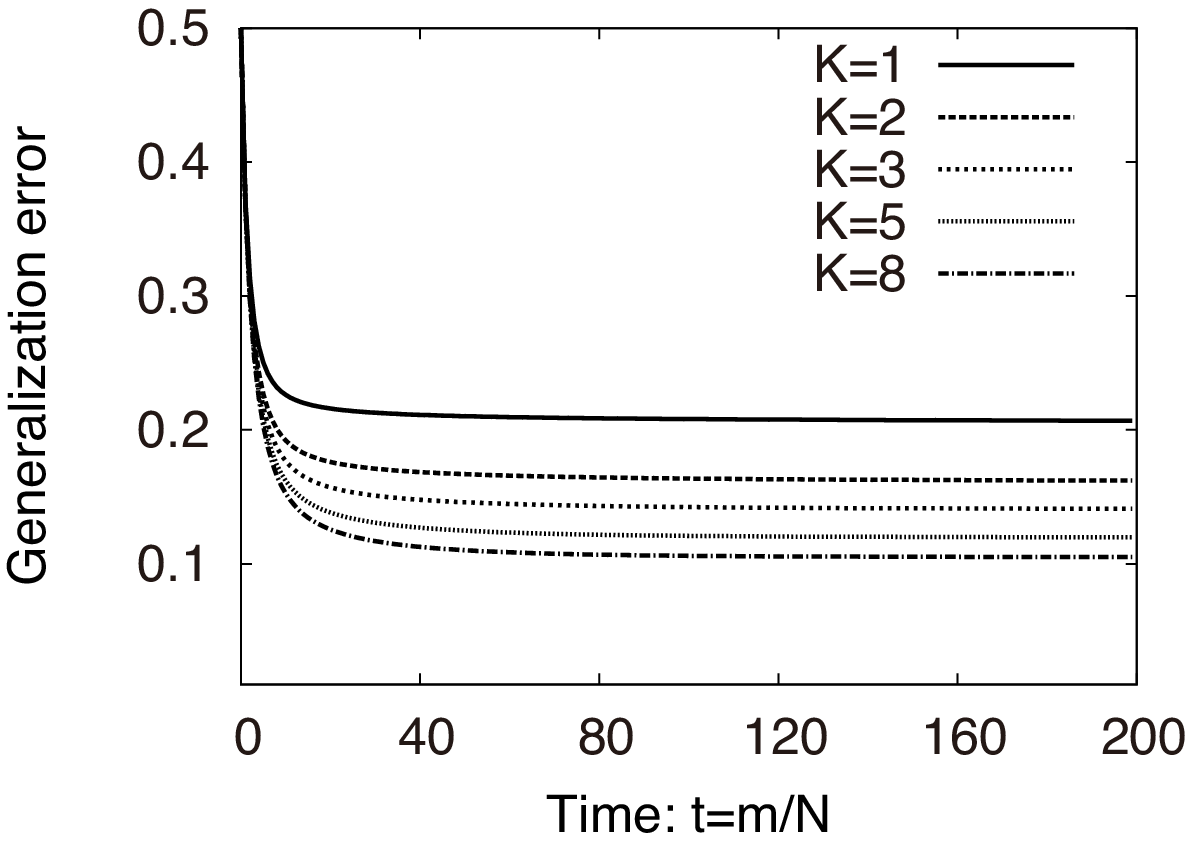}
(d) $\kappa=2$
\end{center}
\end{minipage}
\begin{minipage}{7.5cm}
\begin{center}
\includegraphics[width=7.5cm]{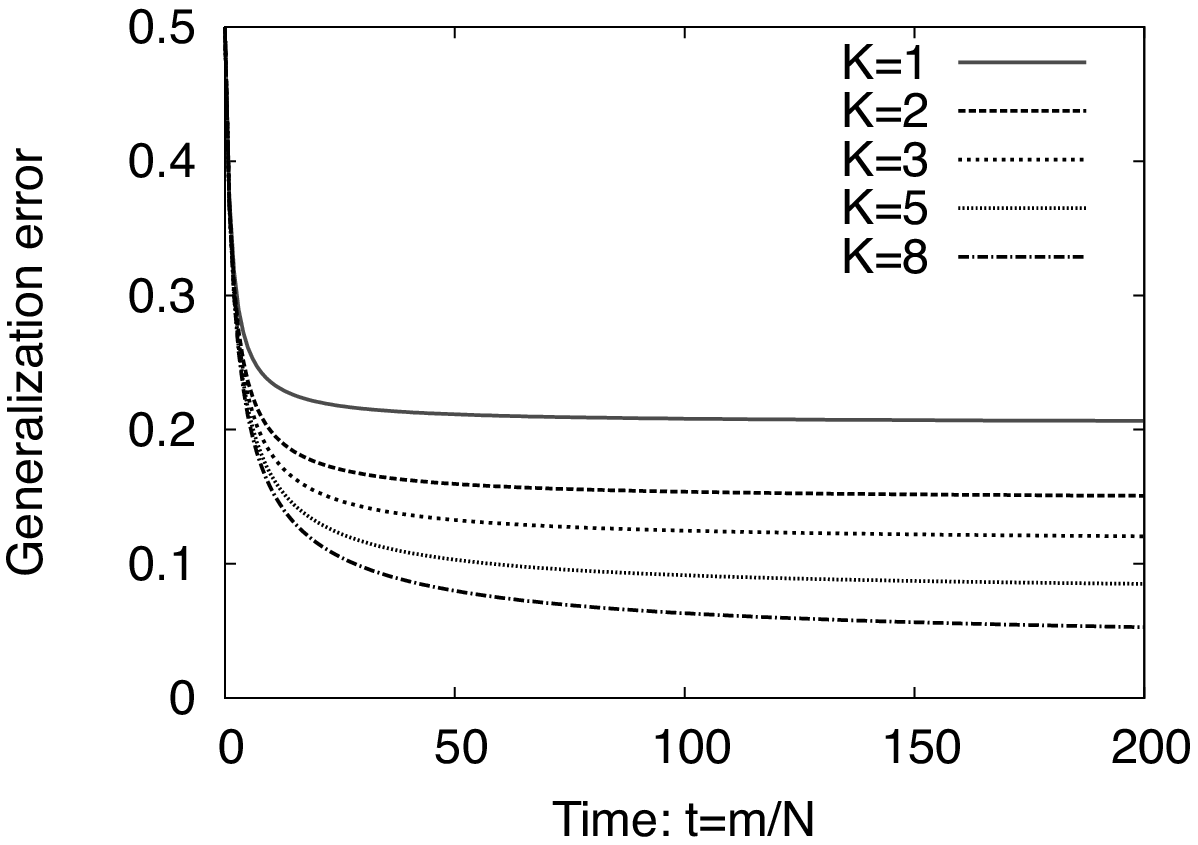}
(e) Hebbian learning rule
\end{center}
\end{minipage}
\caption{\label{eta-and-margin}Time evolution of generalization error for proposed rule. The margin $\kappa$ is (a) $0$, (b) $0.1$, (c) $1$, or (d) $2$. Results obtained using (e) the Hebbian learning rule are shown for comparison. The learning step size is $\eta=1$.}
\end{figure}

Next, we consider the effect of the margin $\kappa$ in the proposed rule.
We will show how the proposed rule changes the learning behavior from the perceptron learning rule to the Hebbian learning rule continuously with increasing size of the margin.
Here, we used analytical solutions.  The learning step size was set to $\eta=1$. 
Figures \ref{eta-and-margin}(a) to \ref{eta-and-margin}(d) show the results for $\kappa=0, 0.1, 1,$ and $2$, respectively. Figure \ref{eta-and-margin}(e) shows the results for the Hebbian learning rule. 
From Fig. \ref{eta-and-margin}(a), the generalization error for the learning process is the same regardless of $K$ when $\kappa=0$. However, in Fig. \ref{eta-and-margin}(b), the effect of the ensemble was increased when using 
a small margin of $\kappa=0.1$.  Figures \ref{eta-and-margin}(c) and \ref{eta-and-margin}(d) show the results obtained using larger margins of $\kappa=1$ and $\kappa=2$, respectively. In these figures, the effect of the ensemble is larger and becomes significant.  In Fig. \ref{eta-and-margin}(e), the Hebbian learning rule gives the most efficient results among  all cases, and are identical to those obtained when $\kappa \rightarrow \infty$. As shown, the perceptron learning rule with a margin is a rule that connects the perceptron learning rule and the Hebbian learning rule continuously through the size of the margin. By setting a margin of $\kappa > 0$, the effect of the ensemble appeared, this effect became significant when a larger margin $\kappa$ was used. 
 
Finally, we clarify the difference between the proposed rule and the perceptron learning rule.
For this purpose, we compare the differential equations for ensemble teacher learning with the perceptron learning rule (eq. (\ref{drj}))  and those for ensemble teacher learning with the perceptron learning rule with a margin (eq. (\ref{drJ-pro})). From eq. (\ref{drj}), $r_j$ remains constant when the overlap between the true teacher and student $R_{J}$ is equal to that between the true teacher and ensemble teachers $R_{B}$. However, according to eq.(\ref{drJ-pro}), $r_j$ is still changing when $R_{J}=R_{B}$ in the proposed rule.  

\section{Conclusion}
In this paper, we proposed novel ensemble teacher learning through the perceptron learning rule with a margin and theoretically analyzed its dynamic behavior. We showed the reason why the perceptron learning rule fails in ensemble teacher learning and that this failure can be avoided by introducing a margin in the perceptron learning rule. Then we derived the order parameter equations of the proposed rule by statistical mechanics methods. The generalization error was obtained from the solutions of the order parameter equations.  The perceptron learning rule with a margin connects the perceptron learning rule and the Hebbian learning rule continuously through the size of the margin. 
Using this rule, we studied the changes in the learning behavior from the perceptron learning rule to the Hebbian learning rule by considering several margin sizes. From the results, we showed that by setting a margin of $\kappa > 0$, the effect of an ensemble appears and becomes significant when a larger margin $\kappa$ is used. Note that the use of a random margin for every learning iteration may also be effective. 

\subsection*{Acknowledgments}
We thank Professor Masato Okada for fruitful discussions. This research was partially supported by the Ministry of Education, Culture, Sports, Science and Technology of Japan through Grant-in-Aid for Scientific Research 21500228.

\appendix

\section{Theory of ensemble teacher learning through the perceptron learning rule}
\label{problem of the perceptron learning rule}
Here, we give theoretical results for ensemble teacher learning through the perceptron learning rule\cite{Utsumi2007}. 
We can use the closed differential equations  (eqs. (\ref{drBkjdt})-(\ref{dldt})) except with $f_k=\Theta(-u v_k) \mbox{sgn}(v_k)$.
$K$ ensemble teachers are used. Utsumi et al. \cite{Utsumi2007} calculated the four averages $\left<f_{k'}v_k\right>$, $\left<f_k y \right>$, $\left<f_k\right>$, and $\left<f_k^2\right>$, then substituted them into eqs. (\ref{drBkjdt})-(\ref{dldt}) to obtain

\begin{align}
\frac{dr_{BJ}}{dt}=&\frac{\eta}{\sqrt{2\pi}} \left(\frac{1+(K-1)q}{K} - R_{BJ}\right) , \label{drbj}\\
\frac{dr_J}{dt}=&\eta \frac{R_{B}-R_{J}}{\sqrt{2\pi}}, \label{drj}\\
\frac{dl}{dt}=&\eta \frac{R_{BJ}-1}{\sqrt{2\pi}} + \eta^2 \frac{1}{\pi} \arccos(R_{BJ}). \label{dl}
\end{align}

\noindent
Here, time $t=m/N$ and we assume $N \rightarrow \infty$. We also assume $R_{BkJ}=R_{BJ}$, $R_{Bk}=R_B$, and $q_{kk'}=q$. $R_B$ and $q$ are constant values. In eqs. (\ref{drbj})-(\ref{dl}), only $dr_{BJ}/dt$ depends on $K$. These solutions remain constant when $R_B=R_J$. This means that the direction cosine $R_J=1$ cannot be achieved when $R_B \neq 1$.
By solving eqs. (\ref{drbj})-(\ref{dl}) numerically at each time $t$, we can obtain the generalization error by substituting $R_J(t)$ into eq. (\ref{generalization_error}).


\begin{thebibliography}{99}
\bibitem{Breiman1996} 
L. Breiman: 
Mach. Learn.  {\bf 24}  (1996) 123.
\bibitem{Freund1997} 
Y. Freund and R. E. Shapire:  
J.  Comput. Syst. Sci.  {\bf 55}  (1997)  119.
\bibitem{Murata2004} 
N. Murata, T. Takenouchi, T. Kanamori, and S. Eguchi:
Neural Comput. {\bf 16} (2004) 1437.
\bibitem{Miyoshi2006}
S. Miyoshi and M. Okada:
J.  Phys.  Soc. Jpn. {\bf 75} (2006) 044002. 
\bibitem{Okada}
M. Okada, K. Hara, and S. Miyoshi: Abstr. Meet. Physical Society of Japan (Spring Meet., 2007) Part 2, p. 319, 21pWB-1 [in Japanese].
\bibitem{Utsumi2007} 
H. Utsumi, S. Miyoshi, and M. Okada: 
J. Phys. Soc. Jpn.  {\bf 76} (2007) 114001. 
\bibitem{Hebb}
D. O. Hebb: {\it The Organization of Behavior} (John Wiley \& Sons., New Jersey, 1949).
\bibitem{Minsky1969}
M. L. Minsky and S. A. Papert: {\it Perceptrons} (The MIT Press, Massachusetts, 1969).
\bibitem{Hara2004} 
K. Hara and M. Okada:
Neural Networks {\bf 17} (2004) 215. 
\bibitem{Hara2011}
K. Hara and S. Miyoshi: {\it Proc. Int. Conf.  Artificial. Neural Net. (ICANN2011)}, European Neural Network Society, Part I, LNCS 6791, (2011)  363.
\bibitem{Hara2009}  K. Hara, Y. Nakayama, S. Miyoshi, and M. Okada:
J. Phys. Soc. Jpn. {\bf 78} (2009) 114001. 
\bibitem{nishimori2001}
H. Nishimori: {\it Statistical Physics of Spin Glass and Information Processing: An Introduction}, (Oxford University Press, Oxford, 2001).
\bibitem{Saadbook}
M. Rattray and D. Saad: in {\it On-line Learning in Neural Networks}, ed. D. Saad (Cambridge University Press, Cambridge, 1998)  183.
\end{thebibliography}
\end{document}